\documentclass[a4paper, oneside, twocolumn, notitlepage, 10pt]{extarticle_ecoc}
\usepackage{ecoc}

\usepackage{color}
\usepackage{xcolor}
\usepackage{graphicx}
\usepackage{tikzscale}
\usepackage{tikz}
\usepackage{pgfplots}
\usetikzlibrary{plotmarks,matrix,chains,scopes,fit,calc,shapes,positioning,decorations,intersections,fit,backgrounds,patterns}
\usetikzlibrary{fadings,shapes.arrows,shadows}
\pgfplotsset{compat=newest} 
\usepackage{pgfplots}
\usepackage{pgffor}
\usepackage{siunitx}
\usepackage{amsmath}
\usepgfplotslibrary[groupplots]
\usetikzlibrary[pgfplots.groupplots]
\usepackage{diagbox}

\pgfplotsset{plot coordinates/math parser=false}
\pgfplotsset{every axis plot/.append style={solid,line width=1.5pt,mark size=1.5pt,mark options={solid,fill=white}}}
\pgfplotsset{every axis legend/.append style={legend cell align=left,font=\footnotesize}}

\usepackage{xparse}
\tikzfading[name=arrowfading, top color=transparent!0, bottom color=transparent!95]
\tikzset{arrowfill/.style={#1,general shadow={fill=black, shadow yshift=-0.8ex, path fading=arrowfading}}}
\tikzset{arrowstyle/.style n args={3}{draw=#2,arrowfill={#3}, single arrow,minimum height=#1, single arrow,
single arrow head extend=.3cm,}}

\NewDocumentCommand{\tikzfancyarrow}{O{2cm} O{FireBrick} O{top color=OrangeRed!20, bottom color=Red} m}{
\tikz[baseline=-0.5ex]\node [arrowstyle={#1}{#2}{#3}] {#4};
} 

\makeatletter
\tikzset{
    block filldraw/.style={
        draw, fill=yellow!20},
    block rect/.style={
        block filldraw, rectangle},
    block/.style={
        block rect, minimum height=0.8cm, minimum width=6em},
    from/.style args={#1 to #2}{
        above right={0cm of #1},
        /utils/exec=\pgfpointdiff
            {\tikz@scan@one@point\pgfutil@firstofone(#1)\relax}
            {\tikz@scan@one@point\pgfutil@firstofone(#2)\relax},
        minimum width/.expanded=\the\pgf@x,
        minimum height/.expanded=\the\pgf@y}}
\makeatother

\begin{document}
\selectlanguage{english}    


\title{Comparison of Single-Wavelength and Multi-Wavelength Transponders in a Physical-layer-aware Network Planning Study}%


\author{
    Jasper M{\"u}ller\textsuperscript{(1,2)}, Ognjen Jovanovic\textsuperscript{(1,3)}, Carmen Mas-Machuca\textsuperscript{(2)}, \\
     Helmut Griesser\textsuperscript{(1)}, Tobias Fehenberger\textsuperscript{(1)}, J{\"o}rg-Peter Elbers\textsuperscript{(1)}
}

\maketitle                  


\begin{strip}
 \begin{author_descr}

   \textsuperscript{(1)} ADVA, Fraunhoferstr. 9a, 82152 Martinsried/Munich, Germany,
   \textcolor{blue}{\uline{jmueller@adva.com}}

   \textsuperscript{(2)} Chair of Communication Networks, Technical University of Munich, Arcisstr. 21, Munich, Germany

   \textsuperscript{(3)} DTU Electro, Technical University of Denmark, 2800 Kgs. Lyngby, Denmark

 \end{author_descr}
\end{strip}

\setstretch{1.1}
\renewcommand\footnotemark{}
\renewcommand\footnoterule{}
\newcommand{\TXOSNR}{OSNR\textsubscript{TX}}
\newcommand{\Pline}{P\textsubscript{line}}

\newcommand{\tf}[1]{\textcolor{red}{#1}}
\newcommand{\jm}[1]{\textcolor{green!60!black}{#1}}
\newcommand{\oj}[1]{\textcolor{blue!50!black}{#1}}
\newcommand{\cmas}[1]{\textcolor{blue}{#1}}


\begin{strip}
  \begin{ecoc_abstract}
    Based on suitable system architectures and realistic specifications, transmit OSNR penalties and spectral constraints of multi-wavelength transponders are identified and analyzed in a network study. We report up to 70\% less required lasers at the expense of a slight increase in number of lightpaths. \textcopyright2022 The Author(s)
  \end{ecoc_abstract}
\end{strip}

\section{Introduction}
\vspace{-6pt}
Optical communication networks have enabled bandwidth-hungry applications for years, and there is no end in sight to the ever-increasing traffic demand. To keep up with these demands in a cost- and power-efficient manner, optical transponder technology has continuously been improved. A key approach to decrease both cost per bit and power consumption has been to scale up the symbol rate per wavelength, with recently announced coherent optical transceivers supporting up to 140~GBd\cite{Jannu}. However, due to the highly challenging requirements of such a large bandwidth, in particular on the electronics, it is not clear how long this scaling of symbol rate remains technically feasible and economically sensible.

A potential way forward is to deploy optical carrier multiplexing, i.e., to use several optical tributary signals on different wavelengths per transponder unit \cite{ITU}. In this case, an integrated multi-wavelength source (MWS) using a single optical power supply, i.e., laser, such as an optical frequency comb can offer significant efficiency improvements over multiple single-wavelength sources (SWS), i.e., lasers, by providing several lines in one integrated component. Significant progress has been made in the MWS subsystem used to generate the lines\cite{kuo2013wideband,gnauck2014comb,anandarajah2015enhanced,zhou201140nm,marin2017microresonator,imran_survey_2018}. MWSs have also been studied on a system\cite{Pfeifle:15,schroder2019laser,Marin-Palomo:20} and architecture\cite{sambo_sliceable_2014} level. 
The implications of using MWSs from a network point of view have also been analyzed, covering novel routing and spectrum allocation algorithms\cite{SantAnna}, provisioning and restoration aspects\cite{dallaglio_impact_2014}, different optical power supply options with respect to techno-economic aspects \cite{imran_techno-economic_2016} and a network throughput study\cite{masood_smart_2020}. 
The impact of MWSs specification in a physical-layer aware network study has not been addressed.

In this paper, we analyze the impact of MWS-based transponders on a network level as to provide guidelines to their specifications and required cost savings. First, suitable architectures for MWS transmitters are described and, depending on practical parameters, realistic transmit optical signal-to-noise ratio (\TXOSNR) values are identified. Using these values along further constraints of MWSs, a network planning study is conducted with two different topologies and varying traffic requests. Comparing MWS with SWS, only a moderate increase in the number of transponders required to fulfill all traffic demands is found, which is mainly due to the lower {\TXOSNR} of MWSs. This penalty is contrasted by potentially significant cost savings and efficiency improvements over SWSs. The presented study gives, to the best of our knowledge for the first time, guidelines on the required MWS specifications and required savings in order for combs to become a viable alternative to SWS transponders in future efficient optical networks.
\vspace{-6pt}
\section{Transmitter Architectures and OSNRs}
\begin{figure}[!t]
\centering
 \includegraphics[width=.9\linewidth]{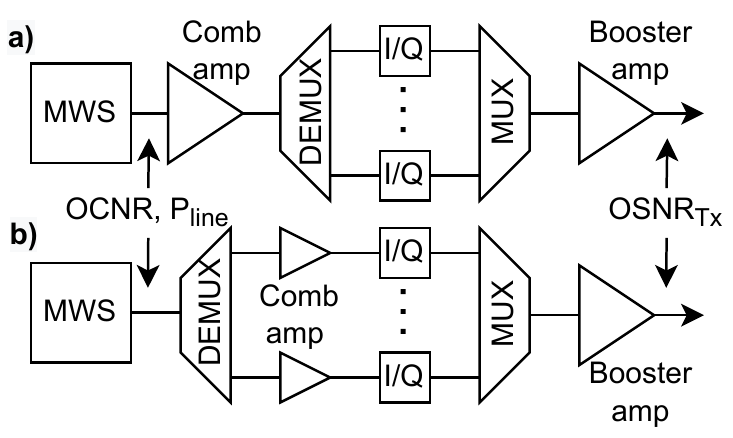}
 \vspace{-0.15cm}
 \caption{Transmitter architecture with \textbf{a)} joint amplification of all comb lines and \textbf{b)} per-line amplification.}
 \label{fig:CombTxArch}
 \vspace{-7pt}
\end{figure}

\vspace{-2pt}
MWSs provide multiple equally spaced optical carriers originating from a single light source. Typically, they are described by their free spectral range (FSR), number of lines, power per line (\Pline), and optical carrier-to-noise ratio (OCNR). Due to multiple lines being generated, optical power and OCNR per line of an MWS are worse than for an SWS. To achieve a sufficiently high power that matches that of an SWS, the MWS lines must be amplified, for which two potential architectures are considered. In Fig.~\ref{fig:CombTxArch}a), an architecture with a single amplifier for all lines is shown. After joint amplification in a comb amplifier (CA), the carriers are separated using a demultiplexer (DEMUX) and each of them is modulated separately with an I/Q modulator. Before launching into the fiber, all carriers are multiplexed and their power is boosted to a launch power of 0~dBm by a booster amplifier (BA). The bottleneck of this architecture is the limited output power (cap) of the CA, which is typically up to 26~dBm \cite{ThorLabsManual}. In the second architecture, shown in Fig.~\ref{fig:CombTxArch}b), this bottleneck is overcome by using a CA for each carrier, resulting into a higher number of required amplifiers. The MUX and DEMUX have a loss of~5~dB\cite{Loss_ref} each, a 5~dB modulation loss is assumed independent of the QAM format\cite{Marin-Palomo:20}, and the modulator insertion and other transmitter losses add up to 23~dB\cite{Loss_ref}. The amplifiers have a 5~dB noise figure.

\begin{figure}[!t]
\centering
  \pgfplotsset{compat=newest}

\pgfplotsset{
    discard if not/.style 2 args={
        x filter/.code={
            \edef\tempa{\thisrow{#1}}
            \edef\tempb{#2}
            \ifx\tempa\tempb
            \else
                \def\pgfmathresult{inf}
            \fi
        }
    }
}

\begin{tikzpicture}[font=\footnotesize]
    \begin{groupplot}[group style={group size=1 by 2, vertical sep=20pt}]
    \nextgroupplot[
        width=\columnwidth,
        height=4cm,
        grid=both,
        ymax = 36.5, ymin = 31.5,
        xmin = -20, xmax = 5,
        ytick distance=1,
        xtick distance=5,
        xlabel shift = -5,
        xlabel=\footnotesize $\text{P}_{\text{line}} \text{ } \si{[dBm]}$,
        ylabel=\footnotesize $\text{OSNR}_{\text{Tx}} \text{ } \si{[dB]}$,
        legend style={font=\normalsize,nodes={scale=0.8, transform shape},fill opacity=0.8, at={(1.01,0.01)}, anchor=south east, draw=none, fill=none},
        legend columns=2
    ]
        \addplot+[discard if not={method}{CW_laser},no marks,color=black,style=dashed] table [x=Pline, y=OSNR, col sep=comma] {figures/OSNR_Tx_Pline.csv};    
        \addlegendentry{SWS}
        \addplot+[discard if not={method}{Comb_arch2},color=red, mark repeat=2,mark phase=2,mark=pentagon*,mark options={fill=.!20!white}] table [x=Pline, y=OSNR, col sep=comma] {figures/OSNR_Tx_Pline.csv};
        \addlegendentry{MWS \textbf{b)}}
        \addplot+[discard if not={method}{Comb_arch1_w_cap},color=green!40!black, mark repeat=2,mark=square*,mark options={fill=.!20!white,solid}] table [x=Pline, y=OSNR, col sep=comma] {figures/OSNR_Tx_Pline.csv};
        \addlegendentry{MWS \textbf{a)} }
    
    \nextgroupplot[
        width=\columnwidth,
        height=4cm,
        grid=both,
        ymax = 36.5, ymin = 31.5,
        xmax = 55, xmin = 33,
        ytick distance=1,
        xtick distance=5,
        xlabel=\footnotesize $\text{OCNR} \text{ } \si{[dB]}$,
        ylabel=\footnotesize $\text{OSNR}_{\text{Tx}} \text{ } \si{[dB]}$,
        xlabel shift = -5,
        legend pos = south east,
        legend style={nodes={scale=0.8, transform shape}},
        legend columns=2
    ]
        
    \addplot+[discard if not={method}{CW_laser},no marks,color=black,style=dashed] table [x=OCNR, y=OSNR, col sep=comma] {figures/OSNR_Tx_OCNR.csv};    
    \addplot+[discard if not={method}{Comb_arch1_w_cap},color=green!40!black, mark repeat=2,mark=square*,mark options={fill=.!20!white,solid}] table [x=OCNR, y=OSNR, col sep=comma] {figures/OSNR_Tx_OCNR.csv};
    \addplot+[discard if not={method}{Comb_arch2},color=red, mark repeat=2,mark phase=2,mark=pentagon*,mark options={fill=.!20!white}] table [x=OCNR, y=OSNR, col sep=comma] {figures/OSNR_Tx_OCNR.csv};
    \end{groupplot}
\end{tikzpicture}

 \vspace{-10pt}
 \caption{{\TXOSNR} vs. power (top) and OCNR (bottom) per line for a MWS with four lines. The MWS curves relate to Fig.~\ref{fig:CombTxArch}.}
 \label{fig:OSNR_OCNR}
\end{figure}
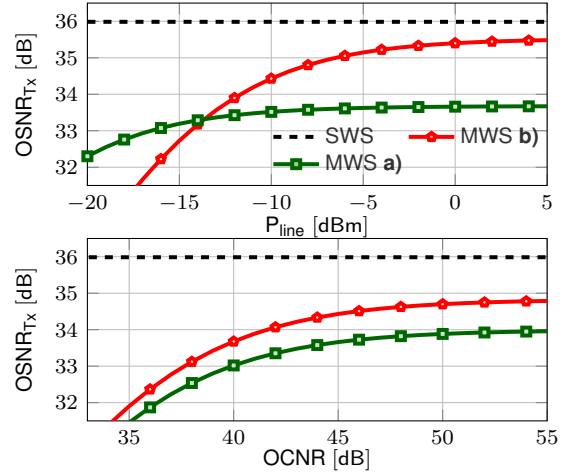

The additional amplifiers and the lower OCNR degrade the {\TXOSNR} of the MWS compared to an SWS. For the given parameters, the reference {\TXOSNR} is around 36 dB for a SWS with OCNR=55~dB and $\text{P}_{\text{line}}$=16~dBm. Using
\vspace{-10pt}
\begin{equation*}
    \text{OSNR}_{\text{Tx}}^{-1} = (\text{OCNR}^{-1} + \text{OSNR}_{\text{CA}}^{-1} + \text{OSNR}_{\text{BA}}^{-1}),
\vspace{-10pt}
\end{equation*}
we investigate the {\TXOSNR} for the two architectures of Fig.~\ref{fig:CombTxArch}. The {\TXOSNR} results for the MWSs in Fig.~\ref{fig:OSNR_OCNR} are shown over per-line power (top) and OCNR (bottom) for typical parameters\cite{HuOxenlowe} of OCNR=45~dB and \Pline=$-\text{10}$~dBm, respectively. We observe that in order not to exceed a 3~dB penalty in {\TXOSNR} compared to an SWS, the power (OCNR) per MWS line must be at least $-\text{14}$~dBm (40~dB), which is achievable by state-of-the-art MWSs \cite{Marin-Palomo:20,HuOxenlowe,schroder2019laser}. In the following, we analyze in a network planning study how these MWS penalties translate into additional transponders. The degrated OCNR of the local oscillator is not considered in this study.

\vspace{-6pt}
\section{Network Planning Study: Setup}
\vspace{-1pt}
For the conducted study, the symbol rates (SRs) and QAM formats are listed in Tab.~\ref{tbl:configuration}. The required SNR for each configuration was obtained by taking the theoretical SNR that achieves the FEC threshold at a bit error rate of 3.5\% for each QAM format as baseline and adding the implementation penalties of Tab.~\ref{tbl:configuration}, which was based on the specifications of commercial transponders.

\begin{table}
\vspace{6pt}
\footnotesize
 \caption{Transponder implementation penalties in dB.}
 \label{tbl:configuration}
 \centering
 \begin{tabular}{|l||c|c|c|c|}
 \hline
 \diagbox[width=10em]{Modulation}{SR [GBd]} & 35 & 70 & 105 & 140 \\
 \hline
 QPSK  & 1 & 1.5 & 2 & 2.5  \\
 \hline
 16QAM  & 1.5 & 2 & 2.5 & 3 \\
 \hline
 64QAM & 2 & 2.5 & 3 & 3.5 \\
 \hline
 \end{tabular}
 \vspace{-0.4cm}
\end{table}


Two network topologies were chosen for the study, representing the different characteristics of a national (Nobel-Germany) and a continental (Nobel-EU) backbone network. The links are assumed to be single bi-directional SSMF links, consisting of 80~km spans with perfect attenuation compensation at the end of each span (EDFA with 5~dB noise figure operating in the C-band). The transmit power spectral density is constant for all SRs. The considered traffic model is based on the number of data centers and internet exchange points in each ROADM location \cite{patri2020planning}. In order to vary the network traffic demands, the individual demands are scaled by the same factor in order to reach different levels of aggregate requested traffic (ART). The routing, configuration and spectrum assignment (RCSA) algorithm considers k=3 shortest-path routing and uses the first-fit algorithm for spectrum assignment. Configurations are chosen in order to minimize the number of required lightpaths (LPs). Only configurations with a required SNR threshold lower than the computed SNR are considered. The SNR takes into account \TXOSNR, linear noise and nonlinearities and is calculated with the closed-form GN model \cite{GN}. For flexible-FSR MWS (flex MWS), we assume that FSRs can be arbitrarily chosen and each line can be routed seperately. Hence, the only difference to SWS is the lower \TXOSNR. The fixed-FSR MWS (fixed MWS) with required co-propagation imposes additional restrictions on the RCSA algorithm, specifically on the spectrum assignment, as spectral slots must be allocated for all MWS lines even if not all lines are in use. All MWSs are assumed to generate four lines to provide a lower bound for MWS-based solutions.

Four different scenarios are investigated for the planning: (1) using only conventional SWS transponders, using only flex MWS transponders, assuming an \TXOSNR~penalty of either (2) 1~dB or (3) 3~dB, and finally (4) planning with fixed MWS transponders requiring co-propagation and SWS transponders. For the last scenario, a fixed FSR of 150~GHz is considered and the MWS is assumed to have an {\TXOSNR} penalty of 1~dB. Also, the RCSA is modified to use SWSs whenever a demand can be met by placing a single LP. Otherwise, a fixed MWS transponder is placed and spectrum is allocated for all lines of the MWS while only the lines needed to meet the traffic demand are activated. 

\begin{figure}[!t]
\centering
  \begin{tikzpicture}[font=\footnotesize]

\definecolor{color0}{rgb}{0.549019607843137,0,0.0588235294117647}
\definecolor{color1}{rgb}{0.0823529411764706,0.690196078431373,0.101960784313725}

\begin{groupplot}[group style={group size=1 by 2, vertical sep=20pt}]
\nextgroupplot[
legend cell align={left},
legend columns=1,
legend style={font=\small, nodes={scale=0.75, transform shape},
  column sep=-1.8,
  draw=none,
  fill=none,
  text opacity=1,
  at={(0.0,1.035)},
  anchor=north west
},
tick align=outside,
tick pos=left,
x grid style={white!69.0196078431373!black},
xmajorgrids,
xmin=10,
xmax=140,
xtick style={color=black},
y grid style={white!69.0196078431373!black},
ylabel={Number of provisioned lightpaths (LPs)},
every axis y label/.append style={at=(ticklabel cs:-0.2)},
ymajorgrids,
ytick style={color=black},
width=0.9\columnwidth,
height=0.5*\columnwidth,
ylabel shift = -6pt,
]
\addlegendimage{thick, red!60!black, mark=*,
                mark options={solid,fill=.!20!white}}
\addlegendentry{SWS}
\addlegendimage{thick, blue!60!black, densely dotted, mark=diamond*,
                mark options={solid,fill=.!20!white}}
\addlegendentry{flex MWS 1dB}
\addlegendimage{thick, orange!40!black, dashed, mark=triangle*,
                mark options={solid,fill=.!20!white}}
\addlegendentry{flex MWS 3dB}
\addlegendimage{thick, black, dash dot, mark=square*,
                mark options={solid,fill=.!20!white}}
\addlegendentry{fix MWS 1dB}

\addplot [thick, red!60!black, mark=*,
                mark options={solid,fill=.!20!white}]
table {%
16.406 128
32.812 141
49.218 152
65.624 174
82.03 192
98.436 212
114.842 225
131.248 246
};
\addplot [thick, blue!60!black, densely dotted, mark=diamond*,
                mark options={solid,fill=.!20!white}]
table {%
16.406 128
32.812 145
49.218 157
65.624 176
82.03 194
98.436 214
114.842 230
131.248 249
};
\addplot [thick, orange!40!black, dashed, mark=triangle*,
                mark options={solid,fill=.!20!white}]
table {%
16.406 130
32.812 150
49.218 161
65.624 183
82.03 200
98.436 227
114.842 241
131.248 260
};

\addplot [thick, black, dash dot, mark=square*,
                mark options={solid,fill=.!20!white}]
table {%
16.406 128
32.812 143
49.218 152
65.624 174
82.03 190
98.436 206
114.842 213
131.248 220
};
\draw[color=green!50!black, very thick](131.2,253) ellipse (3 and 18);
\node[green!50!black,fill=white,opacity=.2,text opacity=1,align=left,font=\scriptsize] at (119,262) {no UP};
\node[green!50!black,fill=white,opacity=.2,text opacity=1,align=left,font=\scriptsize] at (100,150) {UP fixed \\ MWS}; 



\node[inner sep=2pt,fill=white,draw,thin] at (rel axis cs:0.5,0.9) {\textbf{GER}};

\nextgroupplot[
legend cell align={left},
legend style={
  font=\small,
  fill opacity=0.8,
  draw opacity=1,
  text opacity=1,
  at={(0.03,0.97)},
  anchor=north west,
  draw=white!80!black
},
tick align=outside,
tick pos=left,
x grid style={white!69.0196078431373!black},
xlabel={Aggregate requested traffic (ART) [TBit/s]},
xlabel shift = -5,
xmajorgrids,
xtick style={color=black},
y grid style={white!69.0196078431373!black},
ymajorgrids,
ytick style={color=black},
width=0.9\columnwidth,
height=0.5*\columnwidth,
ylabel shift = -5pt,
]
\addplot [thick, red!60!black, mark=*,
                mark options={solid,fill=.!20!white}]
table {%
19.2495 166
28.87425 173
38.499 177
48.12375 193
57.7485 209
67.37325 221
76.998 227
86.62275 231
};
\addplot [thick, blue!60!black, densely dotted, mark=diamond*,
                mark options={solid,fill=.!20!white}]
table {%
19.2495 166
28.87425 174
38.499 178
48.12375 196
57.7485 210
67.37325 221
76.998 226
86.62275 235
};
\addplot [thick, orange!40!black, dashed, mark=triangle*,
                mark options={solid,fill=.!20!white}]
table {%
19.2495 167
28.87425 177
38.499 184
48.12375 197
57.7485 216
67.37325 224
76.998 229
86.62275 239
};

\addplot [thick, black, dash dot, mark=square*,
                mark options={solid,fill=.!20!white}]
table {%
19.2495 166
28.87425 173
38.499 177
48.12375 192
57.7485 207
67.37325 212
76.998 215
86.62275 212
};
\draw[color=green!50!black, very thick](86.5,235) ellipse (2 and 10);
\node[green!50!black,fill=white,opacity=.2,text opacity=1,align=left,font=\scriptsize] at (79,240) {no UP};
\node[green!50!black,fill=white,opacity=.2,text opacity=1,align=left,font=\scriptsize] at (70,175) {UP fixed \\ MWS};
\end{groupplot}

\begin{groupplot}[group style={group size=1 by 2, vertical sep=20pt}]
\nextgroupplot[
axis y line*=right,
ytick pos=right,
xmin=10,
xmax=140,
ylabel={\textcolor{green!50!black}{Underprovisioning (UP) [\%]}},
every axis y label/.append style={at=(ticklabel cs:-0.2)},
ytick style={color=black},
yticklabel style={anchor=west},
tick align=outside,
xtick=\empty, axis line style=transparent,
ymin=0, ymax=100,
ytick style={color=black},
width=0.9\columnwidth,
height=0.5*\columnwidth,
ylabel shift = -6pt,
]

\addplot [thick, green!50!black, dash dot, mark=square*,
                mark options={solid,fill=.!20!white}]
table {%
16.406 0
32.812 0
49.218 0
65.624 0
82.03 0
98.436 4.01
114.842 10.17
131.248 12.42
};

\nextgroupplot[
axis y line*=right,
ytick pos=right,
ytick style={color=black},
yticklabel style={anchor=west},
tick align=outside,
xtick=\empty, axis line style=transparent,
ymin=0, ymax=100,
ytick style={color=black},
width=0.9\columnwidth,
height=0.5*\columnwidth,
ylabel shift = -6pt,
]
\addplot [thick, green!50!black, dash dot, mark=square*,
                mark options={solid,fill=.!20!white}]
table {%
26.3 0
36.2 0
46.7 0
57. 0
66.5 0.66
77.3 1.21
84.9 4.47
95.7 16.82
};
\node[inner sep=2pt,fill=white,draw,thin] at (rel axis cs:0.5,0.9) {\textbf{EU}};
\end{groupplot}

\end{tikzpicture}

 \vspace{-0.4cm}
 \caption{Required number of LPs with respect to provisioned traffic on the Germany (upper) and EU (lower) topology. The dB numbers represent {\TXOSNR} penalties of MWS over SWS.}
 \label{fig:NLP_ART}
 \vspace{-0.15cm}
\end{figure}
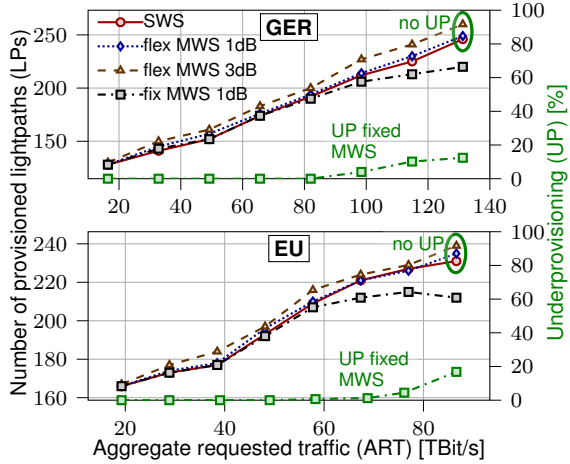

\vspace{-6pt}
\section{Network Planning Study: Results}
Fig.~\ref{fig:NLP_ART} shows the number of required LPs for varying ART for both topologies and all planning scenarios, as well as underprovisioning (UP) for the fixed MWS scenario that describes the ratio of requested data rate that cannot be provisioned to ART. On the Germany topology, the number of required LPs in the flex MWS scenario with 3~dB penalty is up to 7\% higher than for the SWS scenario, whereas this difference is only at most 3\% for the EU network. This is due to a longer average path length of 1100~km on the EU topology compared to 420~km for Germany, which gives a smaller impact of the {\TXOSNR} penalty on the overall SNR. The performance of the fixed MWS scenario is close to SWS in terms of required number of LPs. This scenario, however, is less efficient in terms of the usage of spectrum, as extra slots are allocated for unused MWS lines. Therefore this scenario exhibits UP for high ART levels, as shown in Fig.~\ref{fig:NLP_ART}. As ART increases further, the high spectral occupancy causes the number of deployed LPs to stay approximately constant, showing only small variations due to increases in requested traffic for individual demands and leading to increasing UP. The investigated flex MWS scenarios reduce the number of required optical power supplies by approx.\ 70\% compared to the SWS scenario.
For fixed MWS, due to co-propagation, savings up to 40\% (10\%) on Germany (EU) topology are achieved. Assuming 33\% of the overall transponder cost is the SWS, a four-line flex MWS should cost less than 2.6 times an SWS, to be economically viable.

\begin{figure}[!t]
\centering
\pgfplotsset{
    legend image with text/.style={
        legend image code/.code={%
            \node[anchor=center] at (0.3cm,0cm) {#1};
        }
    },
}
\begin{tikzpicture}[font=\footnotesize]

\begin{axis}[
width=0.9\columnwidth,
tick align=outside,
tick pos=left,
height=0.5\columnwidth, 
x grid style={darkgray176},
xlabel shift = -5pt,
xlabel={{\TXOSNR} penalty [dB]},
xmin=-0.05, xmax=3.05,
xtick style={color=black},
y grid style={darkgray176},
ylabel=\textcolor{red!80!black}{Additional LPs [\%]},
ymin=-0.25, ymax=7.42924528301886,
ytick style={color=black},
ylabel shift = -5pt,
legend columns=2,
legend cell align={left},
legend style={
  font=\small,
  fill opacity=0.8,
  draw opacity=1,
  text opacity=1,
  at={(0.96,-0.1)},
  anchor=north east,
  draw=none,
  fill=none
},
]
\addplot [semithick, red!80!black, mark=triangle*, mark options={solid,fill=.!20!white}]
table {%
0 0
0.1 0
0.2 0
0.3 0
0.4 0
0.5 0
0.6 0
0.7 0
0.8 0
0.9 0.943396226415105
1 0.943396226415105
1.1 0.943396226415105
1.2 0.943396226415105
1.3 0.943396226415105
1.4 0.943396226415105
1.5 0.943396226415105
1.6 2.35849056603774
1.7 3.77358490566038
1.8 3.77358490566038
1.9 3.77358490566038
2 3.77358490566038
2.1 3.77358490566038
2.2 3.77358490566038
2.3 3.77358490566038
2.4 3.77358490566038
2.5 4.24528301886793
2.6 5.18867924528301
2.7 5.66037735849056
2.8 5.66037735849056
2.9 6.13207547169812
3. 7.0754716981132
};
\addplot [semithick, red!80!black, densely dotted, mark=*, mark size=1, mark options={solid,fill=.!20!white}]
table {%
0 0
0.1 0
0.2 0
0.3 0.478468899521522
0.4 0.478468899521522
0.5 0.478468899521522
0.6 0.478468899521522
0.7 0.478468899521522
0.8 0.478468899521522
0.9 0.478468899521522
1 0.478468899521522
1.1 0.478468899521522
1.2 0.956937799043067
1.3 0.956937799043067
1.4 0.956937799043067
1.5 0.956937799043067
1.6 1.43540669856459
1.7 1.43540669856459
1.8 1.43540669856459
1.9 1.43540669856459
2 1.43540669856459
2.1 1.43540669856459
2.2 1.43540669856459
2.3 1.43540669856459
2.4 1.43540669856459
2.5 1.91387559808613
2.6 1.91387559808613
2.7 1.91387559808613
2.8 2.87081339712918
2.9 2.87081339712918
3. 3.34928229665072
};

\node[name=GERbox,inner sep=0pt,outer sep=0pt,anchor=center] at (1.61,4.5) {GER};
\draw[thick,->] (GERbox.south west) -- (1.33,4.);
\draw[thick,->] (GERbox.south east) -- (1.9,4);
\node[name=EUbox,inner sep=0pt,outer sep=0pt,anchor=center] at (1.35,6.55) {EU};
\node[name=EUbox,inner sep=0pt,outer sep=0pt,anchor=center] at (2,0.8) {EU};
\end{axis}

\begin{axis}[
width=0.9\columnwidth,
height=0.5\columnwidth,
axis y line*=right,
tick align=outside,
x grid style={darkgray176},
xmin=-0.05, xmax=3.05,
xtick pos=left,
xtick style={color=black},
y grid style={darkgray176},
ylabel=\textcolor{blue!70!black}{avg SNR diff [dB]},
ytick pos=right,
ytick style={color=black},
yticklabel style={anchor=west},
ylabel shift = -5pt,
legend columns=2,
legend cell align={left},
legend style={
  font=\small,
  fill opacity=0.,
  draw opacity=1,
  text opacity=1,
  at={(0.92,1.02)},
  anchor=north east,
  draw=none,
  fill=none
},
]
\addplot [semithick, blue!70!black, mark=triangle*, mark options={solid,fill=.!20!white}]
table {%
0 0
0.1 0
0.2 -0.050995731277375
0.3 -0.0752777648875735
0.4 -0.102445664787414
0.5 -0.127235160921909
0.6 -0.152362674280269
0.7 -0.17776940665194
0.8 -0.203457061827596
0.9 -0.193451471361403
1 -0.241811678617939
1.1 -0.268238642555911
1.2 -0.294954097936081
1.3 -0.321959804668634
1.4 -0.365226723084298
1.5 -0.397349817966294
1.6 -0.375199124022242
1.7 -0.382744569550542
1.8 -0.414690912167906
1.9 -0.443230025522272
2 -0.476995170317007
2.1 -0.519028098787427
2.2 -0.548581615732392
2.3 -0.578445788080412
2.4 -0.608622401279074
2.5 -0.614544447462464
2.6 -0.642106216694714
2.7 -0.689479865236077
2.8 -0.720641031219742
2.9 -0.757755613885781
3. -0.751860151725413
};
\addplot [semithick, blue!70!black, densely dotted, mark=*, mark size=1, mark options={solid,fill=.!20!white}]
table {%
0 0
0.1 -0.00881495052215442
0.2 -0.0177962452217066
0.3 0.0184202811234968
0.4 0.00911651195691476
0.5 -0.000363003186294009
0.6 -0.0100209444600274
0.7 -0.0198600135784694
0.8 -0.0298829334970314
0.9 -0.040092448083513
1 -0.0504913217796599
1.1 -0.061082339253467
1.2 -0.0642812405209696
1.3 -0.0752409025423049
1.4 -0.0864009189846655
1.5 -0.0977641544154615
1.6 -0.0996588971041419
1.7 -0.135467880872902
1.8 -0.147551274789176
1.9 -0.159848709615719
2 -0.172363089627055
2.1 -0.185097335780643
2.2 -0.198096156307237
2.3 -0.211278847509643
2.4 -0.224690260869551
2.5 -0.228752617749457
2.6 -0.242608244771365
2.7 -0.256701473951832
2.8 -0.28197883543757
2.9 -0.296234132247378
3. -0.302351618690935
}; 

\end{axis}

\end{tikzpicture}

 \vspace{-12pt}
 \caption{Impact of flex-MWS {\TXOSNR} penalty on additionally required LPs (left) and average SNR difference between SWS and MWS transponders (right).}
\label{fig:Txpenalty}
 \vspace{-6pt}
\end{figure}
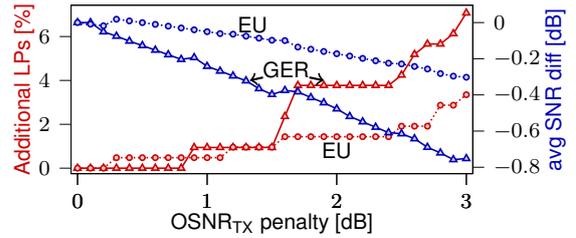

For fixed ART, Fig.~\ref{fig:Txpenalty} shows the impact of  {\TXOSNR} penalty on the required number of LPs as well as the SNR (averaged over all deployed LPs) of MWS minus SWS. We observe that using MWSs in a network study leads to minor drawbacks such as additional LPs for flex MWSs.

\vspace{-6pt}
\section{Conclusions}
In this network planning study, we consider specifications of state-of-the-art MWSs for high-bandwidth transponder configurations. For 4-line flexible MWSs, we show savings of around 70\% in the number of required lasers. In exchange, up to 7\% additional transponders are required for flexible MWSs with 3~dB \TXOSNR~penalty. Fixed MWSs also offer savings in required lasers without requiring additional LPs, but can cause underprovisioning for networks with high spectral occupancy, motivating MWS-aware RCSA algorithms\cite{dallaglio_impact_2014}. DSP benefits\cite{DSP-CPE,lundberg2018frequency,DSP-MIMO} offered by MWSs as well as higher MWS line number are to be treated in future work.

\vspace{-9pt}
\section{Acknowledgements}
\vspace{-1pt}
\footnotesize The work has been partially funded by the German Federal Ministry of Education and Research in the project STARFALL (16KIS1418K) and the 
project 6G-life (16KISK002), as well as the European Research Council through the ERC-CoG FRECOM project (grant no. 771878).
\clearpage

\printbibliography

@article{Marin-Palomo:20,
author = {Pablo Marin-Palomo and Juned N. Kemal and Tobias J. Kippenberg and Wolfgang Freude and Sebastian Randel and Christian Koos},
journal = {Opt. Express},
keywords = {Coherent communications; Frequency combs; Light sources; Nonlinear optical fibers; Optical amplifiers; Signal processing},
number = {9},
pages = {12897--12910},
publisher = {OSA},
title = "{Performance of chip-scale optical frequency comb generators in coherent WDM communications}",
volume = {28},
month = {4},
year = {2020},
%url = {http://opg.optica.org/oe/abstract.cfm?URI=oe-28-9-12897},
doi = {10.1364/OE.380413}
}

@article{HuOxenlowe,
author = {Hao Hu and Leif K. Oxenl{\o}we},
doi = {doi:10.1515/nanoph-2020-0561},
%url = {https://doi.org/10.1515/nanoph-2020-0561},
title = "{Chip-based optical frequency combs for high-capacity optical communications}",
journal = {Nanophotonics},
number = {5},
volume = {10},
year = {2021},
pages = {1367--1385}
}

@online{ThorLabsManual,
  author        = "{ThorLabs}",
  title         = "{EDFA300S and EDFA300P C-Band Erbium-Doped Fiber Amplifiers Manual}",
  year          = "2021",
  url           = "https://www.thorlabs.com/drawings/27a9ab9974814aaa-C16C5F98-D46E-FA32-A941CD90EFBD1216/EDFA300S-Manual.pdf",
}

@online{Jannu,
  author        = "{Acacia}",
  title         = "{Acacia Unveils Industry's First Single Carrier 1.2T Multi-Haul Pluggable Module}",
  year          = "2022",
  url           = "https://acacia-inc.com/blog/acacia-unveils-industrys-first-single-carrier-1-2t-multi-haul-pluggable-module/",
}

@INPROCEEDINGS{patri2020planning,
  author={Patri, Sai K. and Autenrieth, A. and Elbers, J.-P. and Mas-Machuca, C.},
  booktitle={2020 European Conference on Optical Communications (ECOC)}, 
  title={{Planning Optical Networks for Unexpected Traffic Growth}}, 
  year={2020},
  volume={},
  number={},
  pages={1-4},
  doi={10.1109/ECOC48923.2020.9333215}
 }

@article{GN,
author = {Zefreh, Mahdi and Forghieri, Fabrizio and Piciaccia, Stefano and Poggiolini, P.},
year = {2020},
month = {05},
pages = {1-1},
title = {{Accurate closed-form real-time EGN model formula leveraging machine-learning over 8500 thoroughly randomized full C-band systems}},
volume = {PP},
journal = {Journal of Lightwave Technology},
doi = {10.1109/JLT.2020.2997395}
}

@article{SantAnna,
  author={Dallaglio, M. and Giorgetti, A. and Sambo, N. and Velasco, L. and Castoldi, P.},
  journal={Journal of Lightwave Technology}, 
  title="{Routing, Spectrum, and Transponder Assignment in Elastic Optical Networks}", 
  year={2015},
  volume={33},
  number={22},
  pages={4648-4658},
  doi={10.1109/JLT.2015.2477898}}

@article{Loss_ref,
  author={Pillai, Bipin Sankar Gopalakrishna and Sedighi, Behnam and Guan, Kyle and Anthapadmanabhan, N. Prasanth and Shieh, William and Hinton, Kerry J. and Tucker, Rodney S.},
  journal={Journal of Lightwave Technology}, 
  title="{End-to-End Energy Modeling and Analysis of Long-Haul Coherent Transmission Systems}", 
  year={2014},
  volume={32},
  number={18},
  pages={3093-3111},
  doi={10.1109/JLT.2014.2331086}
}

@article{kuo2013wideband,
  title="{Wideband parametric frequency comb as coherent optical carrier}",
  author={Kuo, Bill P-P and Myslivets, Evgeny and Ataie, Vahid and Temprana, Eduardo G and Alic, Nikola and Radic, Stojan},
  journal={Journal of Lightwave Technology},
  volume={31},
  number={21},
  pages={3414--3419},
  year={2013},
  publisher={IEEE}
}

@article{gnauck2014comb,
  title="{Comb-based 16-QAM transmitter spanning the C and L bands}",
  author={Gnauck, Alan H and Kuo, Bill Ping Piu and Myslivets, Evgeny and Jopson, Robert M and Dinu, Mihaela and Simsarian, Jesse E and Winzer, Peter J and Radic, Stojan},
  journal={IEEE Photonics Technology Letters},
  volume={26},
  number={8},
  pages={821--824},
  year={2014},
  publisher={IEEE}
}

@article{anandarajah2015enhanced,
  title="{Enhanced optical comb generation by gain-switching a single-mode semiconductor laser close to its relaxation oscillation frequency}",
  author={Anandarajah, Prince M and D{\'u}ill, Se{\'a}n P {\'O} and Zhou, Rui and Barry, Liam P},
  journal={IEEE Journal of Selected Topics in Quantum Electronics},
  volume={21},
  number={6},
  pages={592--600},
  year={2015},
  publisher={IEEE}
}

@article{zhou201140nm,
  title="{40nm wavelength tunable gain-switched optical comb source}",
  author={Zhou, Rui and Latkowski, Sylwester and O’Carroll, John and Phelan, Richard and Barry, Liam P and Anandarajah, Prince},
  journal={Optics Express},
  volume={19},
  number={26},
  pages={B415--B420},
  year={2011},
  publisher={Optical Society of America}
}

@article{marin2017microresonator,
  title="{Microresonator-based solitons for massively parallel coherent optical communications}",
  author={Marin-Palomo, Pablo and Kemal, Juned N and Karpov, Maxim and Kordts, Arne and Pfeifle, Joerg and Pfeiffer, Martin HP and Trocha, Philipp and Wolf, Stefan and Brasch, Victor and Anderson, Miles H and others},
  journal={Nature},
  volume={546},
  number={7657},
  pages={274--279},
  year={2017},
  publisher={Nature Publishing Group}
}

@article{Pfeifle:15,
author = {Joerg Pfeifle and Vidak Vujicic and Regan T. Watts and Philipp C. Schindler and Claudius Weimann and Rui Zhou and Wolfgang Freude and Liam P. Barry and Christian Koos},
journal = {Opt. Express},
number = {2},
pages = {724--738},
publisher = {OSA},
title = "{Flexible terabit/s Nyquist-WDM super-channels using a gain-switched comb source}",
volume = {23},
month = {1},
year = {2015},
url = {http://opg.optica.org/oe/abstract.cfm?URI=oe-23-2-724},
doi = {10.1364/OE.23.000724},
}

@article{schroder2019laser,
  title="{Laser frequency combs for coherent optical communications}",
  author={Schr{\"o}der, Jochen and F{\"u}l{\"o}p, Attila and Mazur, Mikael and Lundberg, Lars and Helgason, {\'O}skar B and Karlsson, Magnus and Andrekson, Peter A and others},
  journal={Journal of Lightwave Technology},
  volume={37},
  number={7},
  pages={1663--1670},
  year={2019},
  publisher={IEEE}
}

@inproceedings{masood_smart_2020,
	title = {Smart {Provisioning} of {Sliceable} {Bandwidth} {Variable} {Transponders} in {Elastic} {Optical} {Networks}},
	booktitle = {{IEEE} {Conference} on {Network} {Softwarization} ({NetSoft})},
	author = {Masood, M Umar and Khan, Ihtesham and Ahmad, Arsalan and Imran, Muhammad and Curri, Vittorio},
	month = jun,
	year = {2020},
	pages = {85--91},
}

@inproceedings{dallaglio_impact_2014,
	title = {Impact of {SBVTs} based on multi-wavelength source during provisioning and restoration in elastic optical networks},
	doi = {10.1109/ECOC.2014.6963842},
	booktitle = {{ECOC}},
	author = {Dallaglio, M. and Giorgetti, A. and Sambo, N. and Castoldi, P.},
	month = sep,
	year = {2014},
}

@inproceedings{imran_techno-economic_2016,
	title = {Techno-{Economic} {Analysis} of {Carrier} {Sources} in {Slice}-able {Bandwidth} {Variable} {Transponders}},
	booktitle = {{ECOC}},
	author = {Imran, Muhammad and Errico, A. D. and Lord, Andrew and Poti, Luca},
	month = sep,
	year = {2016},
}

@article{sambo_sliceable_2014,
	title = "{Sliceable transponder architecture including multiwavelength source}",
	volume = {6},
	issn = {1943-0639},
	doi = {10.1109/JOCN.2014.6850200},
	number = {7},
	journal = {Journal of Optical Communications and Networking},
	author = {Sambo, Nicola and D'Errico, Antonio and Porzi, Claudio and Vercesi, Valeria and Imran, Muhammad and Cugini, Filippo and Bogoni, Antonella and Potì, Luca and Castoldi, Piero},
	month = jul,
	year = {2014},
	pages = {590--600},
}

@article{imran_survey_2018,
	title = {A {Survey} of {Optical} {Carrier} {Generation} {Techniques} for {Terabit} {Capacity} {Elastic} {Optical} {Networks}},
	volume = {20},
	doi = {10.1109/COMST.2017.2775039},
	number = {1},
	journal = {IEEE Communications Surveys Tutorials},
	author = {Imran, Muhammad and Anandarajah, Prince M. and Kaszubowska-Anandarajah, Aleksandra and Sambo, Nicola and Potí, Luca},
	year = {2018},
	pages = {211--263},
}

@ARTICLE{DSP-MIMO,
  author={Mazur, Mikael and Schröder, Jochen and Karlsson, Magnus and Andrekson, Peter A.},
  journal={Journal of Lightwave Technology}, 
  title="{Joint Superchannel Digital Signal Processing for Effective Inter-Channel Interference Cancellation}", 
  year={2020},
  volume={38},
  number={20},
  pages={5676-5684},
  doi={10.1109/JLT.2020.3001716}
}

@INPROCEEDINGS{DSP-CPE,
  author={Lundberg, Lars and Mazur, Mikael and Lorences-Riesgo, Abel and Karlsson, Magnus and Andrekson, Peter A.},
  booktitle={2017 European Conference on Optical Communication (ECOC)}, 
  title="{Joint Carrier Recovery for DSP Complexity Reduction in Frequency Comb-Based Superchannel Transceivers}", 
  year={2017},
  volume={},
  number={},
  pages={1-3},
  doi={10.1109/ECOC.2017.8346044}
}

@article{lundberg2018frequency,
  title="{Frequency comb-based WDM transmission systems enabling joint signal processing}",
  author={Lundberg, Lars and Karlsson, Magnus and Lorences-Riesgo, Abel and Mazur, Mikael and Torres-Company, Victor and Schr{\"o}der, Jochen and Andrekson, Peter A},
  journal={Applied Sciences},
  volume={8},
  number={5},
  pages={718},
  year={2018},
  publisher={MDPI}
}

@online{ITU,
  author        = "{Recommendation ITU-T G.807}",
  title         = "{Generic functional architecture of optical media network - Amendment 1}",
  year          = "2020",
  url           = "https://www.itu.int/rec/T-REC-G.807-202101-I!Amd1/en",
}

\end{document}